
\hfuzz 2pt
\font\titlefont=cmbx10 scaled\magstep1
\font\bigscript=cmbxsl10 scaled\magstep1
\magnification=\magstep1

\null
\vskip 1.5cm
\centerline{\titlefont DIRECT {\bigscript CP}-VIOLATION}
\medskip
\centerline{\titlefont AS A TEST OF QUANTUM MECHANICS}
\vskip 3.5cm
\centerline{\bf F. Benatti}
\smallskip
\centerline{Dipartimento di Fisica Teorica, Universit\`a di Trieste}
\centerline{Strada Costiera 11, 34014 Trieste, Italy}
\centerline{and}
\centerline{Istituto Nazionale di Fisica Nucleare, Sezione di 
Trieste}
\vskip 1cm
\centerline{\bf R. Floreanini}
\smallskip
\centerline{Istituto Nazionale di Fisica Nucleare, Sezione di 
Trieste}
\centerline{Dipartimento di Fisica Teorica, Universit\`a di Trieste}
\centerline{Strada Costiera 11, 34014 Trieste, Italy}
\vskip 3cm
\centerline{\bf Abstract}
\smallskip

\midinsert
\narrower\narrower\noindent
Direct $CP$-violating effects in the neutral kaon system result in violations
of certain Bell-like inequalities. The new experimental results
on the determination of the phenomenological parameter $\varepsilon'$
allow to dismiss a large class of ``hidden variable'' alternatives
to quantum mechanics.
\endinsert
\bigskip
\vfil\eject

\noindent
{\bf 1. Introduction}
\medskip

The entangled character of neutral kaon pairs
produced at $\phi$-factories offers the possibility of meaningful tests
of Bell's locality [1, 2] in a broad class of hidden-variable 
extensions of ordinary quantum mechanics. This may be achieved in essentially
two different ways, either by exploiting the correlations at 
different times involving strangeness oscillations,[3-9] or by
identifying different kaon states via their
decay products, with or without using regeneration methods.[10-14]

A different, more general point of view is however possible;
as suggested in [15], a clean confirmation of
quantum mechanics against local hidden-variable theories, 
rather than by correlation measures, might come from an accurate 
determination of the phenomenological quantity $\varepsilon'$ by whatever 
means obtained.
The complex parameter $\varepsilon'$ characterizes direct 
$CP$-violation in the neutral kaon system and is
predicted to be non-zero by the Standard Model.[16-18] 

Indeed, as discussed below, inequalities of Clauser-Horne type [19, 20]
can be derived for hidden-variable theories that reproduce neutral 
kaon's phenomenology and comply with the hypothesis of stochastic 
independence of neutral kaon decays.
Then, a specific Bell's inequality is showed to be violated
if $\varepsilon'$ is measured to be non-zero.

A preliminary result on the determination of the phenomenological
quantity ${\cal R}e(\varepsilon'/\varepsilon)$ has been recently announced by 
two experimental collaborations [21, 22] allowing an estimate of the parameter
$\varepsilon'$. We shall explicitly 
show that, as a byproduct, these measurements represent the first experimental
evidence of a violation of Bell's locality occurring in a subnuclear 
system, without requiring correlation measures.
We will make clear that the violation comes about because
of an internal inconsistency of a large class of hidden-variable theories
that pretend to reproduce standard kaon phenomenology
in the presence of direct $CP$-violation.

\vskip 2cm

\noindent
{\bf 2. The neutral kaon system}
\medskip

The standard effective description of neutral kaons makes use
of a two dimensional Hilbert space.[16]
A useful orthonormal basis in this space is given by the $CP$-eigenstates 
$$
|K_1\rangle={|K^0\rangle+|\overline{K^0}\rangle\over\sqrt{2}}\ ,\qquad 
|K_2\rangle={|K^0\rangle-|\overline{K^0}\rangle\over\sqrt{2}}\ ,
\eqno(2.1)
$$
where $|K^0\rangle$, $|\overline{K^0}\rangle$ are the strangeness eigenstates.
The time-evolution and decay of neutral kaons is described by
the nonhermitian phenomenological hamiltonian 
$H_{\rm eff}=M -i{\mit{\Gamma}}/2$, with $M$ and $\mit{\Gamma}$ the positive 
$2\times 2$ mass and decay matrices.
A kaon, initially in a state $|K\rangle$, will evolve up to its proper-time 
$\tau$ into a state $|K(\tau)\rangle=\exp(-i\tau H_{\rm eff})|K\rangle$.

As $CP$-invariance is not preserved in kaon time-evolution, 
the eigenstates of the hamiltonian $H_{\rm eff}$ are two non-orthogonal
admixtures of the states (2.1), which, in the $|K_1\rangle$, 
$|K_2\rangle$ basis, are given by
$$
|K_S\rangle={1\over\sqrt{1+|\epsilon_S|^2}}\pmatrix{1\cr\epsilon_S}
\ ,\quad
|K_L\rangle={1\over\sqrt{1+|\epsilon_L|^2}}\pmatrix{\epsilon_L\cr1}\ .
\eqno(2.2)
$$
The states $|K_S\rangle$, $|K_L\rangle$ correspond to the eigenvalues 
$\lambda_{S,L}\equiv m_{S,L}-i\gamma_{S,L}/2$, where
$m_S$, $m_L$ and $\gamma_S$, $\gamma_L$ are  the masses, 
and widths of the physical short and long-lived kaons.
The complex  quantities $\epsilon_{S,L}$ signal indirect $CP$ and 
$CPT$-violating effects.

The decay-rate, or probability per unit time that a kaon in a generic state
$|K\rangle$ decays into a final state $f$ can be written as
$$
\Gamma(K\to f)=\int{\rm d}\Omega_f\ |{\cal A}(K\to f)|^2\ ,
\eqno(2.3)
$$
where ${\cal A}(K\to f)$ is the Lorentz-invariant amplitude which depends
in general on the so-called Dalitz variables, 
while ${\rm d}\Omega_f$ represents 
the corresponding phase-space measure.

Because of the linearity of the decay-amplitudes with respect to the kaon
states, the effective description allows one to associate the decay
into a specific final state $f$ 
with a $2\times 2$ positive operator $\widetilde{\cal O}_f$. 
In fact, using the decay amplitudes of 
the $CP$-eigenstates $|K_1\rangle$ and $|K_2\rangle$ into $f$, 
one constructs the matrix
$$
\widetilde{\cal O}_{f}\equiv|{\cal A}(K_1\to f)|^2\,
\pmatrix{1&r_f\cr
r_f^*&|r_f|^2}\ ,\quad
r_f={{\cal A}(K_2\to f)\over{\cal A}(K_1\to f)}\ .
\eqno(2.4)
$$
The parameter $r_f$ can be expressed in terms of $\epsilon_{S,L}$ and
the corresponding ratio of amplitudes for the decays of the states
$|K_L\rangle$, $|K_S\rangle$ into $f$; 
these quantities are accessible to the experiment.

In this way, the decay-rate (2.3) can equivalently be written as the
following mean-value
with respect to the kaon state $|K\rangle$:
$$
\Gamma(K\to f)=\int{\rm d}\Omega_f\ \langle K|\widetilde{\cal O}_f|K\rangle\, =
\langle K|\Bigl(\int{\rm d}\Omega_f\,\widetilde{\cal O}_f\,\Bigr)|K\rangle\ .
\eqno(2.5)
$$ 
Since $\widetilde{\cal O}_f$ has zero determinant, it is proportional to
a projector 
$$
{\cal O}_f={1\over 1+|r_f|^2}\
\pmatrix{
1&r_f\cr
r_f^*&|r_f|^2}
=|K_f\rangle\langle K_f|\ ,
\eqno(2.6)
$$
where
$$
|K_f\rangle={1\over\sqrt{1+|r_f|^2}}\left(
\matrix{1\cr r_f^*}\right)\ .
\eqno(2.7)
$$
Further, to any specific final state $f$ one associates an
orthonormal basis in the effective Hilbert space.
In fact,
${\cal O}^\perp_f\equiv 1-{\cal O}_f$ projects
onto the state $|K^\perp_f\rangle$,
$$
{\cal O}_f^\perp=
|K^\perp_f\rangle\langle K^\perp_f|\ ,\qquad
|K^\perp_f\rangle={1\over\sqrt{1+|r_f|^2|}}\left(
\matrix{r_f\cr-1}\right)\ ,
\eqno(2.8)
$$
such that $\langle K_f|K^\perp_f\rangle=0$;
note that $|K^\perp_f\rangle$ cannot decay into 
$f$:
$$
{\cal A}(K^\perp_f\to f)\propto
r_f{\cal A}(K_1\to f)-{\cal A}(K_2\to f)=0\ .
\eqno(2.9)
$$

Since kaons are spinless, in the case of two-body final states $f$ 
the decay amplitudes are constant.
Therefore, 
$$
\Gamma(K\to f)=\langle K|\widetilde{\cal O}_f|K\rangle\, \Omega_f\equiv
|\langle K|K_f\rangle|^2\, {\rm Tr}(\widetilde{\cal O}_f)\, \Omega_f\ ,
\eqno(2.10)
$$
where $\Omega_f$ is just a constant phase-space contribution.
Let us point out that the probability 
$|\langle K|K_f\rangle|^2$ that a kaon in the state $|K\rangle$
be in the state $|K_f\rangle$ is directly related to a decay-rate.
It can actually be measured; indeed, the factor
${\rm Tr}\, (\widetilde{\cal O}_f)$ in (2.10)
can be extracted from the following branching ratios [10] 
$$
BR(K_{S,L}\to f)= {\rm Tr}\
(\widetilde{\cal O}_f)\,{|\langle K_{S,L}|K_f\rangle|^2\over\gamma_{S,L}}
\,\Omega_f\ ,
\eqno(2.11)
$$
where the scalar products $\langle K_{S,L}|K_f\rangle$ are fixed by
equations (2.2) and (2.7) in terms of the parameter $r_f$ in (2.4).

When the final state $f$ comprises more than two particles, the decay
amplitudes are not constant and depend on the appropriate Dalitz variables.
One can still define states $|K_f\rangle$, but only for fixed
kinematical configurations.
Then, the operator $\int{\rm d}\Omega_f\, \widetilde{\cal O}_f$
is clearly not proportional to a projector, but rather to a density matrix:
$$
\rho_f={\int{\rm d}\Omega_f\, \widetilde{\cal O}_f\over
{\rm Tr}\,\Bigl[\int{\rm d}\Omega_f\, \widetilde{\cal O}_f\Bigr]}\ .
\eqno(2.12)
$$
In such cases, the probability that the kaon state $|K\rangle$ be in the mixture
$\rho_f$ is still proportional to the decay-rate of $|K\rangle$ into $f$, but
the proportionality factor, {\it i.e.} the denominator in (2.12), 
cannot be easily related to measurable quantities.

In the sequel, we will deal with two-pion,  $f_{00}=\pi^0\pi^0$, 
$f_{+-}=\pi^+\pi^-$, and semileptonic, 
$f_{\ell^+}=\ell^+\pi^-\nu$, final states. 
In the basis  $|K_1\rangle$, $|K_2\rangle$, a convenient parametrization 
of the corresponding $\widetilde{{\cal O}}_f$ 
in terms of phenomenologically measurable parameters
can be given (for a discussion and more details, see [23, 24]).
Indeed, following the previous considerations, to the pion final states 
we associate the projectors
$$
\eqalignno{
&{\cal O}_{+-}=|K_{+-}\rangle\langle K_{+-}|
={1\over 1+|r_{+-}|^2}\ \pmatrix{1&r_{+-}\cr
                              r_{+-}^*&|r_{+-}|^2\cr}\ , &(2.13a)\cr
&{\cal O}_{00}=|K_{00}\rangle\langle K_{00}|
={1\over 1+|r_{00}|^2}\ \pmatrix{1&r_{00}\cr
                              r_{00}^*&|r_{00}|^2\cr}\ , &(2.13b)
}$$
where the small parameters $r_{+-}$ and $r_{00}$ 
can be written as 
$$
r_{+-}=\varepsilon-\epsilon_L+\varepsilon^\prime\ ,\qquad
r_{00}=\varepsilon-\epsilon_L-2\varepsilon^\prime\ ;
\eqno(2.14)
$$
the two phenomenological quantities $\varepsilon$ and  $\varepsilon'$
signal $CP$ and $CPT$-violating effects that occur directly in the decay
amplitudes and are used to parametrize the following constant amplitude 
ratios [16-18]
$$
\eta_{+-}={{\cal A}(K_L\to\pi^+\pi^-)\over{\cal A}(K_S\to\pi^+\pi^-)}
=\varepsilon+\varepsilon'\ ,\quad
\eta_{00}={{\cal A}(K_L\to\pi^0\pi^0)\over{\cal A}(K_S\to\pi^0\pi^0)}
=\varepsilon-2\varepsilon'\ .
\eqno(2.15)
$$

Concerning the semileptonic decays,  three body final states,
the corresponding projector operators ${\cal O}_\ell$ depend in general
on suitable Dalitz variables.
Nevertheless, one usually takes the approximation of neglecting the lepton 
mass;[16] in this case, one can show that ${\cal O}_\ell$ is constant so
that a state $|K_\ell\rangle$ can be defined and the corresponding probabilities
$|\langle K|K_\ell\rangle|^2$ can be related to measurable quantities, as
explained before.  
However, this approximation is not necessary if one assumes, as we will do, 
the validity of the socalled 
$\Delta S=\Delta Q$ rule;[16, 17]
in this case only $K^0$ and not $\overline{K^0}$ can decay into 
$\ell^+\pi^-\nu$.
This allows us to associate
to the final state $\ell^+\pi^-\nu$ the constant projector
$$
{\cal O}_{\ell^+}=|K_{\ell^+}\rangle\langle K_{\ell^+}|
\equiv|K^0\rangle\langle K^0|=
{1\over2}\ \pmatrix{1&1\cr1&1\cr}\ .
\eqno(2.16)
$$

In the following, we deal with the probabilities $\langle K(\tau)|{\cal
O}_{+-}| K(\tau)\rangle$,
$\langle K(\tau)|{\cal O}_{00}| K(\tau)\rangle$ and
$\langle K(\tau)|{\cal O}_{\ell^+}| K(\tau)\rangle$, that a kaon evolved
up to time $\tau$ into a state $|K(\tau)\rangle$ be in one of the states
$|K_{+-}\rangle$, $|K_{00}\rangle$ and $|K_{\ell^+}\rangle$; 
as shown above, all these
probabilities are expressed in terms of experimentally accessible quantities.

A typical situation in which neutral kaon physics can be studied is offered by
the socalled $\phi$-factories.
In such setups, $\phi$-mesons are copiously produced, which mainly decay into
couple of kaons.
In the case of neutral kaons, because of symmetry reasons, the 
resulting state is entangled in a way that resembles 
the singlet state of two spin $1/2$ particles:
$$
|\Psi\rangle=
N\Bigl(|K_S\rangle\otimes |K_L\rangle - |K_L\rangle\otimes |K_S\rangle\Bigr)\ ,
\eqno(2.17)
$$
where $N$ is a normalization constant.
The two kaons fly apart with opposite momentum (in the $\phi$ rest frame) and
evolve each up to its proper time $\tau_i$, $i=1,2$, according to the
effective hamiltonian $H_{\rm eff}$:
$$
|\Psi\rangle\mapsto|\Psi(\tau_1,\tau_2)\rangle\equiv
\Bigl(e^{-i H_{\rm eff}\tau_1}\otimes e^{-i H_{\rm eff}\tau_2}\Bigr)
|\Psi\rangle\ .
\eqno(2.18)
$$
Let us observe that, setting $\tau_1=\tau_2=\tau$, the entanglement
is preserved by the time-evolution:
$$
|\Psi(\tau)\rangle\equiv
\Bigl( e^{-i H\tau}\otimes e^{-i H\tau}\Bigr)|\Psi\rangle=
 e^{-i\Delta m\tau-\gamma\tau}|\Psi\rangle\ ,
\eqno(2.19)
$$
where $\Delta m=m_L-m_S$ and $\gamma=(\gamma_S+\gamma_L)/2$ mediates the
exponential damping due to the system instability.
The similarity with the standard Einstein-Poldosky-Rosen
setting is apparent; this allows various tests of quantum mechanics to be
performed.[1, 3-9, 25]

\vskip 2cm

\noindent
{\bf 3. Bell-like inequalities  with neutral kaons}
\medskip

Using (2.18) or (2.19), one computes
probabilities related to double-decays and therefore
tests of local hidden-variable theories in the neutral kaon context can be 
discussed. 

Because of (2.19), the probabilities ${\cal P}_\tau(K_a,K_b)$ that one kaon be
in a state $|K_a\rangle$ and the other in a state $|K_b\rangle$, at proper
time $\tau$, are given by
$$
\eqalign{
{\cal P}_\tau(K_a,K_b)&=\langle\Psi(\tau)|
\Bigl({\cal O}_a\otimes{\cal O}_b\Bigr)|\Psi(\tau)\rangle\cr
&= e^{-2\gamma\tau}|N|^2\,
\Big|\langle K_S|K_a\rangle\,\langle K_L|K_b\rangle\,-\,
\langle K_S|K_b\rangle\,\langle K_L|K_a\rangle
\Big|^2\ .}
\eqno(3.1)
$$
In the above expression, the projectors operators ${\cal O}_{a,b}$ are
connected with the kaon states $|K_{a,b}\rangle$ as in (2.6); in a similar way
one can construct the probabilities ${\cal P}_\tau(K_a,K^\perp_b)$,
${\cal P}_\tau(K^\perp_a,K_b)$ and ${\cal P}_\tau(K^\perp_a,K^\perp_b)$, by
using also the orthogonal states $|K^\perp_{a,b}\rangle$ and the
corresponding projectors ${\cal O}_{a,b}^\perp$ (cfr. (2.8)).
Furthermore, by replacing ${\cal O}_a$ (or ${\cal O}_b$) in (3.1) with a
unit matrix, one obtains the probability
${\cal P}_\tau(*,K_b)$ (respectively, ${\cal P}_\tau(K_a,*)$)
that one of the two kaons decays into the final state $f_b$ ($f_a$),
the other being undecayed; then, 
${\cal P}_\tau(*,K_b)=
\langle\Psi(\tau)|1\otimes{\cal O}_b|\Psi(\tau)\rangle
={\cal P}_\tau(K_c,K_b)+{\cal P}_\tau(K^\perp_c,K_b)$, for any kaon state
$|K_c\rangle$.

At each proper time $\tau$, one has a situation which is formally
analogous to the one appearing in standard formulations 
of Bell's inequalities using 
spins.[1, 2, 19, 20]
That is, $|\Psi(\tau)\rangle$ plays the role of the singlet state of 
two spin $1/2$ particles emitted by a source and the projectors, 
${\cal O}_{a,b}$, ${\cal O}^\perp_{a,b}$, are the analog of
spin-polarization operators.
The main difference with respect to the
standard Einstein-Poldosky-Rosen context is that, 
while in the spin case the directions
along which to measure the spin-polarization are freely chosen by the
experimenter, in the neutral kaon case only ``polarization'' directions 
that identify specific decay channels are actually allowed. 
Nevertheless, this restriction does not result in a serious
limitation; indeed, as explained in the Appendix, using regeneration
techniques, many final ``polarizations'' states may be experimentally
reachable.

In any hidden-variable extension of quantum mechanics, 
the description embodied in the state $|\Psi(\tau)\rangle$ in (2.19) is
completed with additional parameters $\lambda$ assigning probabilities
$p^\tau_\lambda(K_a,K_b)$ to the double-decays into final states $f_a$ and
$f_b$.
Also, $\lambda$ fixes the probabilities $p^\tau_\lambda(K_a,*)$ and 
$p^\tau_\lambda(*,K_b)$ associated with single decays at time $\tau$
of one of the two kaons, the other one being undecayed.
The additional parameters generally constitute a statistical ensemble described
by a suitable distribution $\rho(\lambda)$ such that
$\int {\rm d}\lambda\ \rho(\lambda)=1$.
Then, one asks that
the quantum mechanical probabilities (3.1) be reproduced by
integration over $\lambda$; for instance:
$$
{\cal P}_\tau(K_a,K_b)=\int{\rm d}\lambda\ \rho(\lambda)\ 
p^\tau_\lambda(K_a,K_b)\ , \eqno(3.2)
$$
while similar relations hold for 
${\cal P}_\tau(*,K_b)$ and ${\cal P}_\tau(K_a, *)$.
As a consequence, because of the singlet-like character of
$|\Psi(\tau)\rangle$, in any hidden-variable theory
the condition \hbox{${\cal P}_\tau(K_a,K_a)=\,0$} 
should also be satisfied.

Following standard arguments, one also assumes that the 
probabilities $p^\tau_\lambda(K_a,K_b)$ 
fulfill the following Bell's locality request
$$
p^\tau_\lambda(K_a,K_b)=p^\tau_\lambda(K_a,*)\, p^\tau_\lambda(*,K_b)\ .
\eqno(3.3)
$$
At any fixed proper time $\tau$, equality (3.3) amounts to standard
Bell's locality.[1, 2, 19, 20]
In terms of spins, it means that space-like separated 
polarization measurements cannot influence each other so that 
the two events have independent statistics.
In the case of neutral kaons, due to explicit time-dependence and 
instability, one might give up the factorization property
in (3.3) without the need of superluminal transmission of information
at the level of the additional parameters.
It is sufficient to imagine hidden-variable theories that
predetermine the statistics of future decay events at the moment of the 
$\phi$-meson decay.
In the following we shall not consider such theories, but rather assume that
kaon decays are local and stochastically independent events, whence (3.3).
This assumption can be put to experimental test using regeneration techniques;
therefore, hidden-variable theories violating it turn out to be quite unnatural
(see the discussion in the Appendix).

Consequences of the assumption (3.3) can be derived using standard
techniques;[19] in particular, introducing three different
final decay states $f_a$, $f_b$ and $f_c$, associated to the kaon
states $|K_a\rangle$, $|K_b\rangle$ and $|K_c\rangle$, one can prove that
the following inequality must hold:[15]
$$
\Bigl|{\cal P}_\tau(K_a,K_b) - {\cal P}_\tau(K_a,K_c)\Bigr|\ 
\leq\ {\cal P}_\tau(K_c,K_b)\ .
\eqno(3.4)
$$ 
At any fixed proper time $\tau$, the derivation of (3.4) follows 
the one in [26] for the
time-independent case.
However, the same class of inequalities can be deduced from the larger class
of Bell's like inequalities which has been obtained in [10]
via an argument which essentially adapts the derivation of Wigner, 
Belinfante and Holt (see [20], section 3.7).

Let us consider for a moment the standard situation based on
photon-polarization measures, where time plays no role.
Despite its simplicity,
when confronting with actual experiments involving 
coincidence countings, it is 
not inequality (3.4) which is tested.
Due to lack of control of the number of (photon) pairs 
produced that actually impinge on the detectors, 
one is generally forced to use more general inequalities, that take into
account all losses and nonidealities of the experimental 
apparatus.[19, 27]  

The situation is different in the neutral kaon context; there, 
one can actually use
an inequality like (3.4) without further assumptions.
Indeed, tests of an inequality of the form
(3.4) can be performed by measuring the phenomenological parameter 
$\varepsilon'$ in (2.14), (2.15) 
and not directly the probabilities appearing in (3.4).
Indeed, by choosing $K_a=K_{\ell^+}$, $K_b=K_{00}$ and $K_c=K_{+-}$,
an expansion to leading orders in $CP$ and 
$CPT$-violating parameters allows us to express
(3.4) in terms of the difference 
$r_{+-}-r_{00}$, or equivalently, using (2.14), in terms of the constant 
$\varepsilon'$.
Then, the inequality to be fulfilled by any local hidden-variable theory
accounting for neutral kaon phenomenology and satisfying equality (3.3)
reads:
$$
\big|{\cal R}e(\varepsilon')\big|\leq 3\, |\varepsilon'|^2\ .
\eqno(3.5)
$$

Typically, $CP$-violations in decay processes are ignored
for sake of simplicity in almost all discussions 
of Bell's inequalities concerning neutral kaons
(for an exception see [14]).
This means $r_{+-}=r_{00}=0$ and in such a case (3.5) reduces to a trivial 
identity.
Indeed, setting $r_{+-}=r_{00}=0$, the projectors 
in (2.13) both coincide with the projector onto the kaon state $|K_1\rangle$, 
the decay of $|K_2\rangle$ into a two-pion final state being forbidden.
With these simplifying assumptions,
there is an identification of the kaon states with definite
strangeness or $CP$-quantum numbers via their decay products.
However, in this way,  one of the polarization-like direction would be
lost. Instead, besides the strangeness eigenstates 
$|K^0\rangle$, $|\overline{K^0}\rangle$,
and the $CP$-eigenstates $|K_1\rangle$, $|K_2\rangle$, a third orthonormal basis
of kaon states $|\widetilde{K}_S\rangle$, $|\widetilde{K}_L\rangle$ is needed
to make the typical Bell's argument run.
In [10], kaon regeneration methods have been proposed to supply it.
The idea of using slabs of regenerating materials in asymmetric
$\phi$-factories has been put forward in [12] and 
reconsidered for other purposes in [13]. 
In the approach of [15]
there is no need to consider regeneration as a tool to
provide a triple of polarization-like directions $K_a$, $K_b$ and $K_c$.
Moreover, it is the very fact that small $CP$-violating
effects are not neglected which allows for a connection between
direct $CP$-violations and violations of Bell's locality.

The parameter $\varepsilon'$ is accessible to experiments and
any experimental determination of it is in principle
able to disprove (3.5) and thus (3.4).
In the next paragraph we consider the results of the KTeV and the NA48
Collaborations which used uncorrelated kaons.[21, 22]  
However, the presence and role of $\varepsilon'$ could in principle be
tested via measuring correlations, that is by determining probabilities of
joint events as those appearing in (3.4).
In fact, as already explained, when the kaon states $|K_{a,b}\rangle$ are
associated to actually occurring decays, then the probabilities 
${\cal P}_\tau(K_a,K_b)$ in (3.1)
are not only formally well-defined, but also measurable quantities.

Finally, let us point out that a violation 
of (3.5) does not provide a test of the 
theoretical framework in which the parameter $\varepsilon'$ can
be estimated.[16-18] 
It only says that any local hidden-variable theory accounting for stochastic 
independence of
kaon decays and reproducing kaon phenomenology and thus the right value of
$\varepsilon'$ must, at the same time, fulfill inequality (3.4).
Therefore, that very
same theory cannot be compatible with a value of $\varepsilon'$ 
which violates (3.5).
However, as any test, also the one based on the inequality (3.5) 
is significant 
only when giving a negative result, saying that an inequality 
of the form (3.4) is violated.
Only in this case, one is able to experimentally exclude
a large set of local deterministic extensions of quantum mechanics.
If the experimental data had turned out to be compatible with
a vanishing value for $\varepsilon'$ (result predicted by the
so-called superweak phenomenological model [28]), 
then, an inequality of the form (3.4) would have simply been unsuited
to exclude the hidden-variable theories considered in this paper.

\vskip 2cm

\noindent
{\bf 4. Experimental results}
\medskip

While the phenomenological parameter $\varepsilon$ is very 
well known and of the order $10^{-3}$,~[29]
the parameter $\varepsilon'$ has only recently been determined with sufficient
accuracy.
With a fixed-target setup that uses uncorrelated kaons, the KTeV
and NA48 Collaborations have measured the double ratio of decay rates in (2.15)
$$
{|\eta_{+-}|^2\over|\eta_{00}|^2}\simeq1+6\,
{\cal R}e\,\left({\varepsilon'\over\varepsilon}\right)\ .
\eqno(4.1)
$$
Their results are  ${\cal R}e\,(\varepsilon'/\varepsilon)=(2.80\pm0.41)\times
10^{-3}$,[21] and ${\cal R}e\,(\varepsilon'/\varepsilon)=(1.85\pm0.73)\times
10^{-3}$,[22]
which are in rough agreement with the theoretical 
predictions.[18]
The Standard Model further predicts the phase $\varphi'$ of 
$\varepsilon'$ to be very close to the phase $\varphi$ of $\varepsilon$.[17]
Assuming then
${\cal R}e\,(\varepsilon'/\varepsilon)=|\varepsilon'/\varepsilon|$, 
these experimental determinations allow 
us to check inequality (3.5), that can be rewritten as
$$
R\equiv\left|{\varepsilon'\over\varepsilon}\right|\left(\,
{\cos\varphi\over|\varepsilon|} -
3\,\left|{\varepsilon'\over\varepsilon}\right|\right)\,\leq\,0\ .
\eqno(4.2)
$$ 
Using the most recent determination for $\varepsilon$ 
and the two previously quoted
experimental results, one finds 
$$
\eqalignno{
&R_{\rm KTeV}=0.89\pm0.13\ , &(4.3a)\cr
&R_{\rm NA48}=0.59\pm0.23\ . &(4.3b)}
$$
The inequality (3.5) is therefore violated by a few standard deviations.
The accuracy of the determination of ${\cal R}e(\varepsilon'/\varepsilon)$
by the two collaborations will be further improved when all the experimental 
data are elaborated (the figures previously quoted 
refer to the study of only about $20\%$ (in the case of KTeV)
and $15\%$ (for NA48) of the collected data).
Correspondingly, the test on the violation of the inequality (3.5) will also
improve; accuracy of about twenty standard deviations can easily be expected.
In this way, the test of Bell's locality using neutral kaons will turn out to be
at the same level of accuracy of the best tests performed with photon
cascades.[30] 

We point out that another kaon experiment, KLOE in Frascati, 
is presently collecting data and its results on
the value of $\varepsilon'/\varepsilon$ will be announced in the near future.
This experiment is particularly interesting for our considerations; it
takes place at the Daphne $\phi$-factory and therefore uses correlated kaons.
Thanks to the machine high luminosity, both 
${\cal R}e(\varepsilon'/\varepsilon)$ and 
${\cal I}m(\varepsilon'/\varepsilon)$ can be measured independently,[31]
so that the working assumption $\varphi'=\varphi$ used above will no longer be
necessary.
Further, as stressed before, the KLOE setup can also perform a direct check of
inequality (3.5) by actually measuring the various probabilities involved.
However, due to efficiency limitations%
\footnote{$^{\dagger}$}{For these reasons, the
actual determination of $\varepsilon'$ at a $\phi$-factory involves measures of
observables that differ from those that enter the probabilities in (3.4).}
(see the discussion in [10]) the accuracy of such a test is expected to be much 
worse than the one using the measure of $\varepsilon'$.

\vskip 2cm

\noindent
{\bf 5. Discussion}
\medskip

As already mentioned, for experimental reasons, the class of
inequalities holding in local hidden-variable theories that are  
tested are in general more complicated than (3.4).[20]
One may say that (3.4) stays on the same level as the 
inequality originally provided by Bell [1] to support the 
incompatibility of local hidden-variable theories
with quantum mechanics.
In the case of neutral kaons, a time-dependent version of it 
reads:
$$
\Bigl|E_\tau(K_a,K_b) - E_\tau(K_a,K_c)\Bigr|\leq 1 + 
E_\tau(K_c,K_b)\ ,
\eqno(5.1)
$$
where the correlation functions can be expressed as
$$
E_\tau(K_a,K_b)\equiv
{\cal P}_\tau(K_a,K_b)\, +\, {\cal P}_\tau(K^\perp_a,K^\perp_b)
\, -\, {\cal P}_\tau(K^\perp_a,K_b)\, -\, {\cal P}_\tau(K_a,K^\perp_b)
\ . \eqno(5.2)
$$
When $\tau=0$, the inequality (5.1) is the standard one and can be clearly
contradicted by quantum mechanics.
None the less, it has not been submitted to any test:
due to actual experimental inefficiencies, one
cannot count on perfect anticorrelation among
pairs in a singlet state as predicted by quantum mechanics
and therefore more complicated inequalities are in
general needed.[20]
Furthermore, for $\tau\neq0$, the factorized common exponential 
damping factor $\exp(-2\gamma\tau)$ in (3.1) compared 
with the constant $1$ on the right hand side
of (5.1), confines the possibility of violations to too
short times after the $\phi$-meson decay.%
\footnote{$^{\dagger}$}{This kind of exponential suppression of
violation of Bell's inequalities also appears in the analysis of [7-9],
where, however, correlations resulting in
having or not having a $\overline{K^0}$ at different proper times 
are used.}

Like (5.1), inequalities of the type (3.4), 
if testable through coincidence counting
only, would also be plagued by all possible kind of apparatus
inefficiencies.
Nevertheless, standard neutral kaon phenomenology 
allows to pass from (3.4) to (3.5) and,
therefore, to a condition on the parameter $\varepsilon'$. 
Since it pertains to
the kaon physics, it must be reproduced by any theory aiming to replace
standard quantum mechanics neutral kaon phenomenology. 
Then, the original inequality can be tested by determining $\varepsilon'$:
no further assumptions are necessary and the corresponding
loopholes are avoided.

Let us stress that the violation of (3.5) comes about because
local hidden-variable theories that reproduce neutral kaon phenomenology and
account for the stochastic independence of neutral kaon decays,
become self-contradictory.
In fact, they must reproduce a physically measurable parameter 
$\varepsilon'$ whose value happens to violate certain inequalities 
which must also be satisfied by the same theories.
Therefore, beside its role in showing the direct violation of the $CP$-symmetry,
the determination of $\varepsilon'/\varepsilon$ at the same time
provides evidence against a whole class of local completions 
of quantum mechanics by showing an internal inconsistency of these theories.

Moreover, the result is the first obtained in a subnuclear system
and without correlation measurements.
Notice, however, that the inequality (3.4) involves joint kaon decays and, 
as discussed, no
apriori obstruction exists to testing it by counting simultaneous decays into
two-pion and semileptonic final states.
Nevertheless, these measures must eventually reproduce the figure of 
$\varepsilon'$ obtained by any other experiment and cannot show a 
confirmation of (3.5).

Obviously, not all local hidden-variable theories can be excluded by
a determination of $\varepsilon'$, but only those for which	  
the inequality (3.4) can be derived.
For this, two requirements need to be satisfied:
$a)$ 
the truth of the quantum mechanical phenomenological description of neutral 
kaons and thus the impossibility of observing simultaneous decays of
singlet-like entangled kaons into the same final states; and 
$b)$
that the underlying local hidden-variable theories do not
predetermine {\it ab initio} future decay events which are thus
stochastically independent.

We remark that request $a)$ is just the acceptance of kaon phenomenology
(ultimately of the Standard Model) as a 
correct description of reality.
Instead, request $b)$ apriori eliminates those local hidden-variable theories
where the probabilities of certain decays are fixed 
at the moment of the $\phi$-meson decay
(models have been proposed in [3-7]). 
Concretely, if we abandoned request $b)$ above, then two-pion 
and semileptonic decays considered in this paper 
might be correlated {\it ab initio}
and we would not be allowed to use the factorization (3.3).
In the standard situation, the experimenter circumvents the possibility that
experimental outcomes be predetermined by freely choosing
which polarization to measure, without entangled photons ``knowing'' it.  
Instead, in the neutral kaon case this is not possible and must be
excluded apriori.

Clearly, having put constraints on the class of local hidden-variable
theories to be tested,
the question is how large is the class which fulfills them.
In other words, one may ask how restrictive is request $b)$ or
how feasible is a hidden-variable theory which
reproduces kaon phenomenology and predetermines the statistics of future 
decay events.
In the Appendix we argue that kaon regeneration 
phenomena [10, 12] may be used to put most severe constraints on them.
Indeed, it is shown that the probabilities in (3.4) might be measured not by
the statistics of 
double-decay events at time $\tau$, but rather at later different times
$\tau+\tau_a$ and $\tau+\tau_b$ after interposition of regenerating material in
one or both the flight paths of the two kaons.
The time $\tau$ can be chosen such that the presence or not
of a slab of regenerating material
cannot be known at the moment of the $\phi$-meson decay.
Furthermore, a lot of freedom is left to the experimenter, since
more than one slab of regeneration material can be inserted across
the kaons path.
Such hidden variable theories, where the decay statistics is
preestablished, should also account for all these 
possible interactions of neutral 
kaons with the regenerators: they clearly turn out to be very
{\it had hoc} and unviable.

\vskip 2cm

\noindent
{\bf Appendix}
\medskip

Because of their different strangeness quantum number and their strong 
interactions with nucleons, the neutral kaons $|K^0\rangle$ and
$|\overline{K^0}\rangle$ acquire different phases while passing 
through slabs of materials
as copper, lead or carbon.[10]
This leads a long-lived kaon $|K_L\rangle$ entering the slab to end up with a
(regenerated) short-lived component $|K_S\rangle$ after its exit.  
Their time-evolution in matter is thus different from the one in vacuum which is
described by the hamiltonian $H_{\rm eff}=M-i{\mit\Gamma}/2$.

A slab of homogeneous regenerating material can be described by a 
regeneration parameter
$$
\displaystyle\rho={\pi\nu\over m_K}{f-\overline{f}\over\lambda_S-\lambda_L}\ ,
\eqno(A.1)
$$
where $\nu$ is the number of scattering centers per unit volume, $m_K$ 
the kaon mass, $f$,
$\overline{f}$ the forward scattering amplitudes 
for $|K^0\rangle$ and $|\overline{K^0}\rangle$
on nucleons and $\lambda_{S,L}$ the eigenvalues of the vacuum effective
hamiltonian.
The time-evolution of neutral kaons in matter can be effectively described by
an effective hamiltonian $H'_{\rm eff}$; in the $|K^0\rangle$, 
$|\overline{K^0}\rangle$ basis it reads
$$
H'_{\rm eff}=H_{\rm eff}-{2\pi\nu\over m_K}\,\pmatrix{f&0\cr
0&\overline{f}}\ .
\eqno(A.2)
$$
For slabs of materials of sufficiently small thickness $d$, the time-evolution 
of the
$H_{\rm eff}$-eigenstates $|K_{S,L}\rangle$ can be
approximated by [10]
$$
\eqalignno{
& e^{-i H'_{\rm eff}\Delta\tau'}|K_S\rangle=|K_S\rangle\, -\,
\, \xi\, |K_L\rangle&(A.3a)\cr
& e^{-i H'_{\rm eff}\Delta\tau'}|K_L\rangle=|K_L\rangle\, +\,
\xi\, |K_S\rangle\ ,&(A.3b)}
$$
where 
$$
\xi=i{\pi\nu\over p_K}(f-\overline{f})\, d\ ,
\eqno(A.4)
$$
with $p_K$ the kaon momentum in the laboratory frame, and $\Delta\tau'$ is
the time spent by the kaons in the material.
  
Consider now two kaons generated in a
$\phi$-meson decay that evolve in vacuum up to equal proper 
times $\tau$, when they enter two slabs of regenerating materials
with regeneration parameters $\rho_a$, $\rho_b$.
After times $\Delta\tau'_a$ and
$\Delta\tau'_b$ they exit and evolve in vacuum  
up to proper times $\tau+\Delta\tau'_a+\Delta\tau_a$ and 
$\tau+\Delta\tau'_b+\Delta\tau_b$. 
Let us set $\tau_a=\Delta\tau'_a+\Delta\tau_a$ and
$\tau_b=\Delta\tau'_b+\Delta\tau_b$, then
the initial correlated state $\Psi$ in (2.17) evolves into
(compare with (2.18))
$$
|\Psi'(\tau+\tau_a,\tau+\tau_b)\rangle\equiv
\left( e^{-iH_{\rm eff}\Delta\tau_a} e^{-iH'_{\rm eff}\Delta\tau'_a}
 e^{-iH_{\rm eff}\tau}\right)\otimes
\left( e^{-iH_{\rm eff}\Delta\tau_b} e^{-iH'_{\rm eff}\Delta\tau'_b}
 e^{-iH_{\rm eff}\tau}\right)|\Psi\rangle\ .
\eqno(A.5)
$$
The probability that the kaon states at proper times $\tau+\tau_a$ and
$\tau+\tau_b$ be two generic states
$|K_{\eta_a}\rangle$ and $|K_{\eta_b}\rangle$, is given by
(see (3.1)):
$$
\eqalign{
&{\cal P}'_{\tau_a,\tau_b}(K_{\eta_a},K_{\eta_b})=
|\langle\Psi'(\tau+\tau_a,\tau+\tau_b)|{\cal O}_{\eta_a}\otimes
{\cal O}_{\eta_b}
|\Psi'(\tau+\tau_a,\tau+\tau_b)\rangle|^2\cr
&= e^{-2\gamma\tau}|N|^2\,\Bigl|
\langle K_{\eta_a}|W(\tau_a)|K_S\rangle
\langle K_{\eta_b}|W(\tau_b)|K_L\rangle\, -\,
\langle K_{\eta_a}|W(\tau_a)|K_L\rangle
\langle K_{\eta_b}|W(\tau_b)|K_S\rangle
\Bigr|^2\ ,
}\eqno(A.6)
$$
where 
$$
\eqalignno{
&W(\tau_a)=e^{-iH_{\rm eff}\Delta\tau_a}
\ e^{-iH'_{\rm eff}\Delta\tau'_a}\ , &(A.7a)\cr
&W(\tau_b)=e^{-iH_{\rm eff}\Delta\tau_b}\
e^{-iH'_{\rm eff}\Delta\tau'_b}\ . &(A.7b)}
$$
Expanding the states $|K_S\rangle$ and $|K_L\rangle$ 
in $\langle K_{\eta_j}|W(\tau_j)|K_{L,S}\rangle$, $j=a,b$, above, 
along orthonormal basis 
$|K_j\rangle$, $|K^\perp_j\rangle$ such that
$$
\langle K_{\eta_j}|W(\tau_j)|K_j^\perp\rangle=0\ ,
\eqno(A.8)
$$ 
one sees that $(A.6)$ becomes
$$
{\cal P}'_{\tau_a,\tau_b}(K_{\eta_a},K_{\eta_b})=
|\langle K_{\eta_a}|W(\tau_a)|K_a\rangle|^2\,
|\langle K_{\eta_b}|W(\tau_b)|K_b\rangle|^2\ 
{\cal P}_\tau(K_a,K_b)\ ,
\eqno(A.9)
$$
with ${\cal P}_\tau(K_a,K_b)$ as in (3.1).
Therefore, at least in principle, the probabilities appearing in (3.4)
can be measured via $(A.9)$ by freely choosing to insert or not a 
regeneration slab along the kaons path, without the kaons
``knowing'' it.

The above argument clearly relies on the condition $(A.8)$.
In order to show that this condition is actually implementable, 
we work in the
$|K_{1,2}\rangle$ basis, where 
$|K_{\eta_j}\rangle\propto(1,\eta_j^*)$ and 
$|K^\perp_j\rangle\propto(r_j,-1)$ as
in (2.7) and (2.8).
Then, neglecting higher powers in the small parameters 
$\epsilon_{S,L}$ and $\xi$, all of order $10^{-3}$ or smaller, 
using $(A.3)$,
after some algebra equality $(A.8)$ leads to the following expression for
the parameter $\xi$ in $(A.4)$:
$$
\xi={(r_j+\epsilon_L+\eta_jr_j\epsilon_S)\,
 e^{(i\Delta m-\Delta\gamma/2)\Delta\tau_j}
-(\eta_j+\eta_j\epsilon_S+\epsilon_L)
\over
 e^{(i\Delta m-\Delta\gamma/2)\Delta\tau_j}
+\eta_jr_j} \ ,
\eqno(A.10)
$$
where $\Delta m=m_L-m_S$ and $\Delta\gamma=\gamma_S-\gamma_L\simeq2\Delta m$.

When $|K_j\rangle=|K_{\ell^+}\rangle$ as in (2.16), $r_j=1$ and no choice of
$\eta_j$ renders the right hand side of $(A.10)$ 
compatible with the smallness of $\xi$.
On the other hand, 
when $|K_j\rangle$ equals $|K_{+-}\rangle$ or $|K_{00}\rangle$,
$r_j$ equals $r_{+-}$ or $r_{00}$ in (2.13); choosing 
$\eta_j=r_{+-}$ or $\eta_j=r_{00}$ and 
neglecting higher orders in the small parameters, $(A.10)$ further simplifies:
$$
\xi=\varepsilon\Bigl(1-\,
e^{(1-i)\Delta m\Delta\tau_j}\Bigr)\ .
\eqno(A.11)
$$
Such an equality is implementable by an appropriately chosen regenerating
material.

Thus, there are no apriori obstructions to
computing  the joint-probabilities ${\cal P}_\tau(K_a,K_b)$
in (3.4) by using $(A.9)$ and thus
double-decay rates at times $\tau+\tau_a$ and $\tau+\tau_b$;
only in the case when one of the kaon states, say $|K_a\rangle$, equals 
$|K_{\ell^+}\rangle$, corresponding to semileptonic 
final states, regeneration techniques fail to provide the link given by
equation $(A.9)$.
In these cases, no regenerating slab is inserted and $\tau_a$ is set to zero.
Then, the challenge of hidden-variable theories to be able to predetermine the
joint-probabilities appearing in (3.4) comes from the fact that we can decide
to obtain the latter by freely choosing  to insert or not
the appropriate regenerating slabs.
For instance, in the simplest situation described above, there are sixteen
available possibilities:
four choices for ${\cal P}_\tau(K_{+-},K_{00})$ and two 
choices for ${\cal P}_\tau(K_{\ell^+},K_{+-})$ and 
${\cal P}_\tau(K_{00},K_{\ell^+})$ that involve semileptonic final states.
Furthermore, one is in principle allowed to insert more than one regenerating
slab in the beam path of either kaons.
In this way, a lot of freedom of choice is left to the experimenter and 
these hidden-variable theories must then be able to account for it; 
they must be very special indeed.

\vfill\break

\vfill\eject

\centerline{\bf REFERENCES}
\bigskip\bigskip

\item{1.} J. Bell, Physics {\bf 1} (1969) 480;
{\it Speakable and Unspeakable in Quantum Mechanics},
(Cambridge University Press, Cambridge, 1987)
\smallskip
\item{2.} M. Redhead, {\it Incompleteness, Nonlocality and Realism},
(Clarendon Press, Oxford, 1987)
\smallskip
\item{3.} F. Selleri, Lett. Nuovo Cim. {\bf 36} (1983) 521
\smallskip
\item{4.} J. Six, Phys. Lett. {\bf A150} (1990) 243
\smallskip
\item{5.} D. Home and F. Selleri, J. Phys. {\bf A24} (1991) L1073
\smallskip
\item{6.} P. Privitera and F. Selleri, Phys. Lett. {\bf B286} (1992) 261
\smallskip
\item{7.} A. Datta and D. Home, Found. Phys. Lett. {\bf 4} (1991) 165
\smallskip
\item{8.} G. Ghirardi, R. Grassi and T. Weber, in the Proceedings
of the {\it Workshop on Physics and Detectors for DA${\mit\Phi}$NE},
G. Pancheri, ed., (INFN-LNF, Frascati, 1991)
\smallskip
\item{9.} G. Ghirardi, R. Grassi and R. Ragazzon, in The Da$\Phi$ne Physics
handbook, Vol. I, L. Maiani, G. Pancheri and N. Paver eds., (INFN, Frascati, 
1992)
\smallskip
\item{10.} A. Di Domenico, Nucl. Phys. {\bf B450} (1995) 293
\smallskip
\item{11.} A. Di Domenico, in {\it Workshop on K Physics},
L. Iconomidou-Fayard, ed., (Editions Frontieres, Gif-sur-Yvette (France),
1997)
\smallskip
\item{12.} P.H. Eberhard, Nucl. Phys. {\bf B398} (1993) 155
\smallskip
\item{13.} A. Bramon and M. Nowakowski, Phys. Rev. Lett. {\bf 83} (1999) 1;
Phys. Rev. D {\bf 60} (1999) 094008
\smallskip
\item{14.} F. Uchiyama, Phys. Lett. {\bf A231} (1997) 295
\smallskip
\item{15.} F. Benatti and R. Floreanini, Phys. Rev. D {\bf 57} (1998) 1332
\smallskip
\item{16.} T.D. Lee and C.S. Wu, Ann. Rev. Nucl. Sci. {\bf 16} (1966) 511
\smallskip
\item{17.} L. Maiani, in {\it The Second Da$\,{\mit\Phi}$ne Physics Handbook}, 
L. Maiani, G. Pancheri and N. Paver, eds., (INFN, Frascati, 1995)
\smallskip
\item{18.} For recent reviews, see: S. Bertolini, M. Fabbrichesi and J.O. Eeg,
Estimating $\varepsilon'/\varepsilon$, {\tt hep-ph/9802405},
Rev. Mod. Phys. {\bf 72}, to appear; A. Buras, Theoretical status of
$\varepsilon'/\varepsilon$, {\tt hep-ph/9908395},
to appear in the {\it Proceedings of Kaon'99}, (Chicago University Press,
Chicago, 1999); see also the recent textbook:
G.C. Branco, L. Lavoura, J.P. Silva, {\it CP violation},
(Oxford University Press, Oxford, 1999)
\smallskip
\item{19.} J.F. Clauser and M.A. Horne, Phys. Rev. D {\bf 10} (1974) 526
\smallskip
\item{20.} J.F. Clauser and A. Shimony, Rep. Prog. Phys. {\bf 41} (1978) 1881
\smallskip
\item{21.} The KTeV Collaboration, Phys. Rev. Lett. {\bf 88} (1999) 22
\smallskip
\item{22.} The NA48 Collaboration, Phys. Lett. {\bf B465} (1999) 335
\smallskip
\item{23.} F. Benatti and R. Floreanini, Phys. Lett. {\bf B401} (1997) 337
\smallskip
\item{24.} F. Benatti and R. Floreanini, Nucl. Phys. {\bf B 511} (1998) 550
\smallskip
\item{25.} R.A. Bertlmann, W. Grimus and B.C. Hiesmayr, 
Phys. Rev. D {\bf 60} (1999) 114032
\smallskip
\item{26.} A. Garuccio, Phys. Rev. A {\bf 52} (1995) 2535
\smallskip
\item{27.} P.H. Eberhard, Phys. Rev. A {\bf 47} (1993) R747
\smallskip
\item{28.} L. Wolfenstein, Phys. Rev. Lett. {\bf 13} (1964) 562
\smallskip
\item{29.} Particle Data Group, Eur. Phys. J. {\bf C3} (1998) 1
\smallskip
\item{30.} G. Weihs, T. Jennewien, C. Simon, H. Weinfurter, A. Zeilinger,
Phys. Rev Lett. {\bf 81} (1998)  5039
\smallskip
\item{31.} G. D'Ambrosio, G. Isidori and A. Pugliese, $CP$ and $CPT$
measurements at Da$\Phi$ne, in 
{\it The Second Da$\,\mit\Phi$ne Physics Handbook}, 
L. Maiani, G. Pancheri and N. Paver, eds., (INFN, Frascati, 1995)

\bye